# Predicting the outcomes of fuel drop impact on heated surfaces using SPH simulation

Xiufeng Yang, Yaoyu Pan, Song-Charng Kong[*]
Department of Mechanical Engineering, Iowa State University, Ames, IA 50011, USA
*Corresponding author: kong@iastate.edu

**Abstract**
The impact of liquid drops on a heated solid surface is of great importance in many engineering applications. This paper describes the simulation of the drop-wall interaction using the smoothed particle hydrodynamics (SPH) method. The SPH method is a Lagrangian mesh-free method that can be used to solve the fluid equations. A vaporization model based on the SPH formulation was also developed and implemented. A parametric study was conducted to characterize the effects of impact velocity and wall temperature on the impact outcome. The present numerical method was able to predict different outcomes, such as deposition, splash, breakup, and rebound (i.e., Leidenfrost phenomenon). The present numerical method was used to construct a regime diagram for describing the impact of an iso-octane drop on a heated surface at various Weber numbers and wall temperatures.

**Introduction**

Drop impact on heat walls is encountered in many engineering applications, such as engine sprays, liquid cooling, and material processing [1-5]. The ability to predict the outcome of drop-wall interactions is critical in helping improve the performance of related engineering processes. For example, the outcome of drop-wall interactions will affect the fuel-air mixture distribution and the subsequent combustion quality.

There have been numerous studies on drop-wall interactions. Sikalo et al. [6] experimentally investigated the effects of impact parameters on the outcome of drop impact on horizontal surfaces. Yeong et al. [7] conducted an experiment to study the impact and rebound dynamics of a drop on a nanocomposite super-hydrophobic surface. Moita and Moreira [8] also characterized the mechanisms of disintegration of water and fuel drops impacting onto heated surfaces. Xu et al. [9] experimentally investigated the effect of ambient gas pressure on drop splashing on a smooth dry surface. Liang et al. [10] studied the boiling phenomenon near the Leidenfrost point of a drop. Staat et al. [11] constructed a regime diagram of ethanol drop impact on a smooth surface at different Weber numbers and wall temperatures.

Different numerical methods have also been proposed to predict drop-wall interactions, e.g., the marker-and-cell method by Harlow and Shannon [12]. Bussmann et al. [13] simulated the fingering and splashing of a drop impacting on a solid surface using a modified volume-of-fluid (VOF) method. Many previous numerical studies have not considered the effect of wall temperature on drop impact. The wall temperature can affect the result significantly, particularly when the wall temperature is higher than the Leidenfrost point. A ghost-fluid level-set method was used by Villegas et al. [14] to simulate the drop impact on a hot surface at a temperature above the Leidenfrost point. In their work, only the rebound of a drop was studied for Weber number lower than 60. In engineering CFD applications, proper models are required to predict the outcome of drop-wall interactions [15, 16]. Such models often use formulas and criteria based on experimental findings at a limited range of operating conditions [17-19].

The purpose of this work is to characterize the details of drop-wall interactions using the smoothed particle hydrodynamics (SPH) method. It is hoped that more accurate drop-wall interaction models can be derived based on such detailed numerical simulations in the near future. In our previous work [20, 21], the SPH method was shown to be able to simulate the process of drop impact on dry or wet surfaces. This work will extend the previous capability by considering the effects of ambient gas conditions, wall temperature, and vaporization on drop-wall interactions.

**Numerical Method**

A more detailed documentation of the present SPH method can be found in previous publications [20-22]. The following is a brief description. Considering the effect of vaporization at the liquid-gas interface, the following governing equations are used.

$$\frac{d\rho}{dt} = -\rho \nabla \cdot \boldsymbol{u} + \dot{m}''' \qquad (1)$$

$$\frac{d\boldsymbol{u}}{dt} = \boldsymbol{g} - \frac{1}{\rho}\nabla p + \frac{\mu}{\rho}\nabla^2 \boldsymbol{u} \tag{2}$$

$$\frac{dT}{dt} = \frac{1}{\rho C_p}\nabla \cdot (\kappa \nabla T) - \frac{h_v}{\rho C_p}\dot{m}''' \tag{3}$$

$$\frac{dY}{dt} = \frac{\nabla \cdot (\rho D \nabla Y)}{\rho} \tag{4}$$

Here $\rho$ is the fluid density, $\boldsymbol{u}$ is the fluid velocity, $\dot{m}$ is the mass evaporation rate across the interface while $\dot{m}'''$ is the volumetric mass evaporation rate, $p$ is the fluid pressure, $\mu$ is the dynamic viscosity, $T$ is the temperature, $C_p$ is the specific heat at constant pressure, $\kappa$ is the thermal conductivity, and $\boldsymbol{g}$ is the gravitational acceleration, $h_v$ is the latent heat of vaporization, $Y$ is the vapor mass fraction, and $D$ is the mass diffusivity of the vapor.

The volumetric mass flux can be calculated as

$$\dot{m}''' = \frac{\dot{m}}{V} = \frac{\nabla \cdot (\rho D \nabla Y)}{1-Y} \tag{5}$$

The following equation of state is used to calculate pressure

$$p = c^2(\rho - \rho_r) + p_r \tag{6}$$

where $c$ is a numerical speed of sound, $\rho_r$ is a reference density and $p_r$ is a reference pressure.

In the SPH method, a continuous fluid is discretized into a set of SPH particles. These particles also have physical properties, such as mass m, density $\rho$, velocity $\boldsymbol{u}$, and viscosity $\mu$. The values of a function $f(\boldsymbol{r})$ and its derivative at point $\boldsymbol{r}_a$ can be approximated using the following particle summation.

$$f(\boldsymbol{r}_a) \approx \sum_b f(\boldsymbol{r}_b) W(\boldsymbol{r}_a - \boldsymbol{r}_b, h) \frac{m_b}{\rho_b} \tag{7}$$

$$\nabla f_a = \sum_b \frac{m_b}{\rho_b} f_b \nabla_a W_{ab} \tag{8}$$

where $W$ is a kernel function. The parameter $h$ is referred to as a smoothing length, which determines the size of the integral domain. $\nabla_a W_{ab}$ denotes the gradient of $W$ taken with respect to the coordinates of particle $a$. By applying the particle summation, the governing equations can be replaced by SPH equations. For more details, please refer to [22].

**Results and Discussion**

The numerical method is first tested to simulate the evaporation of a static n-heptane drop surrounded by high-temperature nitrogen. Then it is applied to study the impact of iso-octane drop on a solid surface at different temperatures, with air as the ambient gas. The physical properties of the liquids and gases are given in Table 1.

Table 1. Physical properties of the liquids and gases

| | $\rho$ (kg/m³) | $\mu$ (kg/m/s) | $\kappa$ (W/m/K) | $C_p$ (J/kg/K) | $M$ (kg/mol) | $h_v$ (J/kg) | $T_B$ (K) | $D_v$ (m²/s) |
|---|---|---|---|---|---|---|---|---|
| Nitrogen | 1.25 | $3\times10^{-5}$ | 0.026 | 1040 | 0.028 | | | $1\times10^{-5}$ |
| n-Heptane | 684 | $4\times10^{-4}$ | 0.12 | 2220 | 0.01 | $3.3\times10^5$ | 372 | |
| Air | 1.2 | $2\times10^{-5}$ | 0.046 | 1000 | 0.029 | | | $9\times10^{-6}$ |
| iso-Octane | 692 | $4\times10^{-3}$ | 0.1 | 2100 | 0.114 | $3.1\times10^5$ | 372 | |

**Evaporation of a static drop**

The initial radius of the drop ($R$) is 0.3 mm with initial temperature 310 K. The computational domain is a circle with a radius of 1.5 mm. The initial temperature of the gas is 648 K. The initial vapor mass fraction in the gas phase is zero. The temperature and vapor mass fraction of the boundary do not change during the simulation. Due to evaporation, the number of liquid particles will decrease while the number of gas particles will increase during the simulation.

Figure 1 shows that the vapor mass fraction is maximum at the liquid-gas interface and decreases with an increase in distance from the interface. As time goes by, heating occurs and the temperature increases; the vapor mass fraction also increases in the gas phase. Figure 2 shows the temperature profiles from the center of the drop to the domain boundary at different times. Although there is a temperature jump at the liquid-gas interface initially, the temperature profile quickly becomes continuously because of heat transfer. Then the drop temperature increases from 310 K to about 340 K shortly. Figures 1 and 2 also show that the size of the drop decreases with time due to the evaporation.

The change of the drop size is depicted in Figure 3, showing the normalized square of drop radius as a function of time. It can be seen that at the early stage the evaporation rate increases slightly as the liquid temperature increases. After 4 seconds, the evaporation rate becomes nearly linear, consistent with the D-square law of drop evaporation [23].

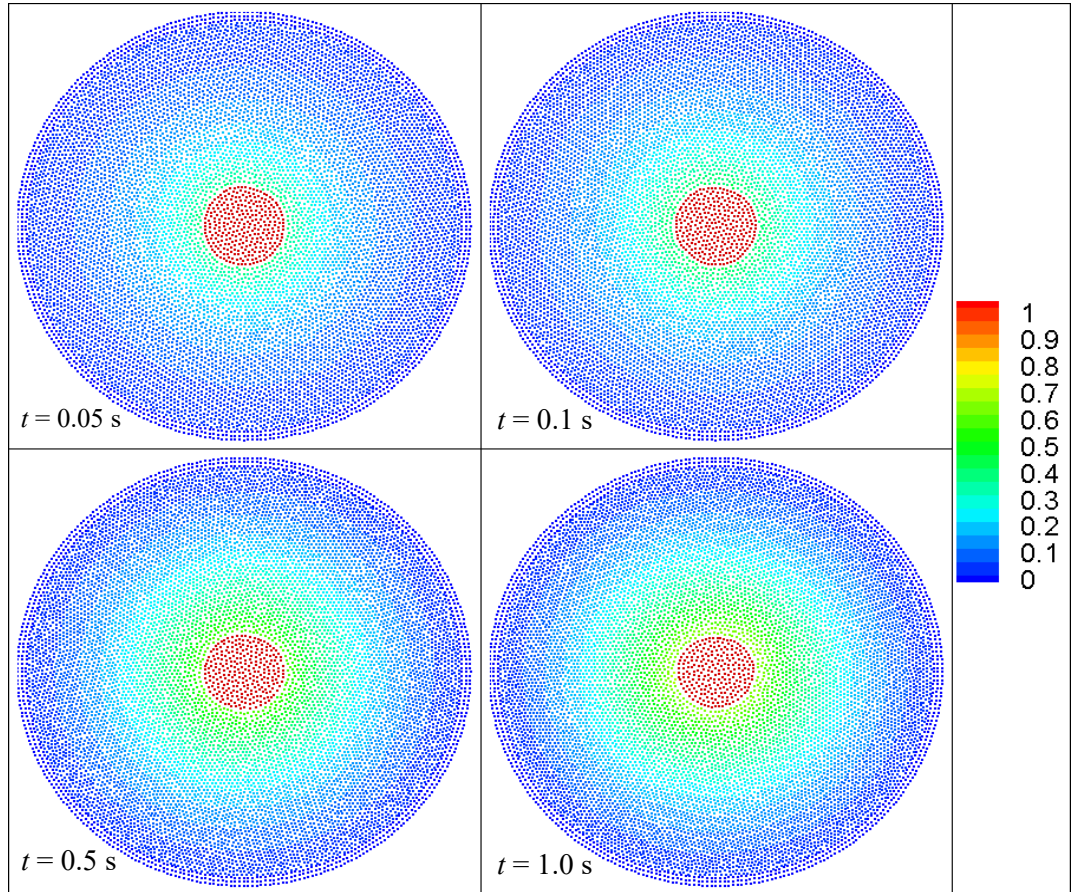

**Figure** 1. Mass fraction of n-heptane at different times.

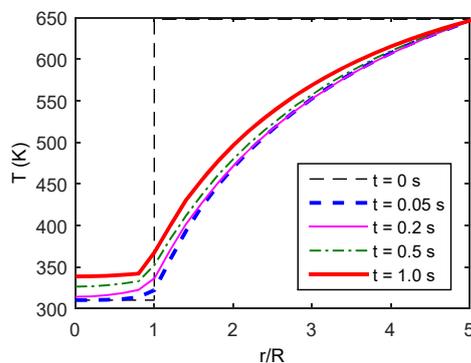

**Figure** 2. Temperature profiles at different times.

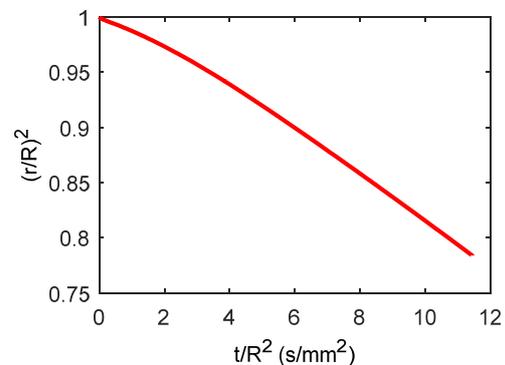

**Figure** 3. Normalized square of drop radius as a function of time.

**Drop impact on a heated surface**

The impact of an iso-octane drop on a heated surface is also simulated. The ambient gas is air. The initial radius of the drop is 25 μm. The initial velocity of the drop varies from 1 to 50 m/s. The initial temperature of the drop is 323 K. The temperature of the solid surface varies from 323 to 523 K and does not change during the simulations.

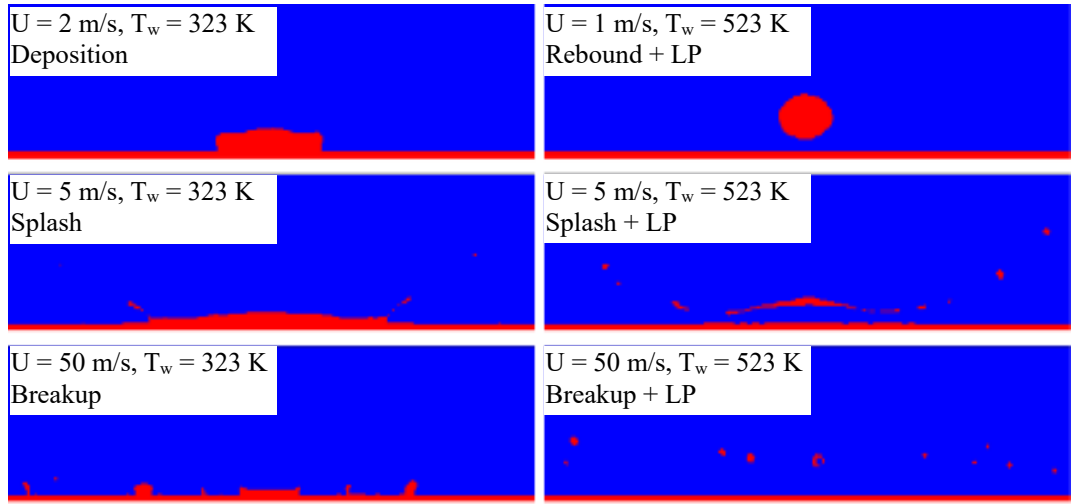

Fig. 4. Different outcomes of drop impact on heated surfaces. "LP" denotes Leidenfrost phenomenon.

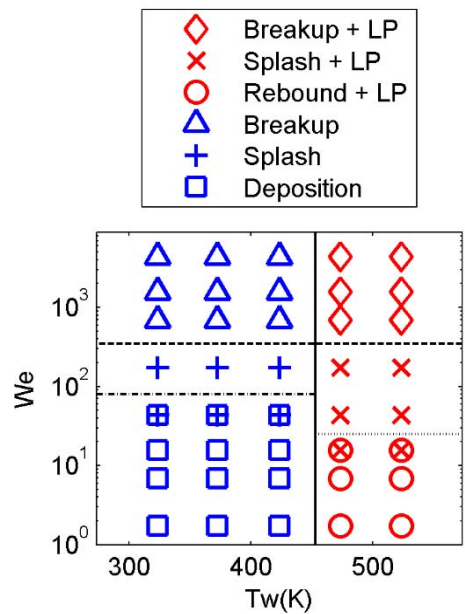

Fig. 5. Impact regimes of drop impact on heated surfaces. The vertical solid line is the Leidenfrost temperature, the dash lines is the critical Weber number of transition from one regime to the other.

The outcomes of drop-wall interactions change with surface temperatures and impact velocities. Six different types of outcomes are identified from the numerical simulations, namely, deposition, splash, breakup, rebound + LP, splash + LP, and breakup + LP, as shown in Figs. 4 and 5. "LP" denotes Leidenfrost phenomenon. For deposition, the drop stays or spreads on the wall without producing splashing liquid. For splash, the drop spreads on the wall and forms a film with secondary droplets generated near the rim of the film. For breakup, the drop breaks up into many smaller droplets on the wall. As the wall temperature increases, more vapor is generated. When the wall temperature is higher than the Leidenfrost temperature, significant vapor is generated between the liquid and wall, pushing the liquid away from the wall. This prediction is consistent with the so-called Leidenfrost phenomenon. The right column of Fig. 4 shows the simulated Leidenfrost phenomenon at different impact velocities. At low velocity, the drop rebounds from the hot wall. At higher velocity, the drop splashes on the wall, and the liquid is also pushed up by the vapor. At much higher velocity, the drop breaks up on the wall and the secondary droplets are also pushed up by the vapor. Fig. 5 shows that the formation of secondary droplets strongly depends on the Weber number, while the drop is easier to splash when the wall temperature is higher than the Leidenfrost temperature. When the wall temperature is lower than the Leidenfrost temperature, the results are similar at different wall temperatures although the evaporation rate is different at different temperatures. Figure 5 is similar to the phase diagram from Staat et al. [11] although the liquids studied are different.

**Summary and Conclusions**

The present study simulates the evaporation of a static drop and the impact of a drop on a heated surface at different temperatures using the SPH method. In simulating drop evaporation, the predicted drop size history is consistent with the traditional D-square law. For drop-wall interactions, three different impact regimes are identified for wall temperature lower than the Leidenfrost point, namely, deposition, splash, and breakup. Similarly, three regimes at wall temperatures above the Leidenfrost point are also predicted, namely, rebound, splash, and breakup. Whether the drop splash or breakup depends on the impact Weber number.

**Acknowledgements**

The authors acknowledge the support of Ford Research and Innovation Center.